# $\beta$-phase $(Al_xGa_{1-x})_2O_3$ thin film with Al composition more than 70%

Che-Hao Liao, Kuang-Hui Li, Carlos G. Torres-Castanedo, Guozheng Zhang, and Xiaohang Li*

King Abdullah University of Science and Technology (KAUST), Advanced Semiconductor Laboratory, Thuwal, 23955-6900, Saudi Arabia

**Keywords:** $\beta$-phase, $(AlGa)_2O_3$; $Ga_2O_3$; diffusion; thermal annealing

**Abstract:**

In this work, we have demonstrated wide-composition-range $\beta$-$(Al_xGa_{1-x})_2O_3$ thin films with record-high Al compositions up to 77% for $\beta$-$(Al_xGa_{1-x})_2O_3$ covering bandgaps from 4.9 to 6.4 eV. With optimized thermal annealing conditions, the $\beta$-$Ga_2O_3$ binary thin films on sapphire substrates transformed to the $\beta$-$(AlGa)_2O_3$ ternary thin films with different compositions. The binary to ternary transformation resulted from the Al atom diffusion from sapphire into the oxide layers; meanwhile, the Ga atoms diffused into sapphire leading to thicker thin films than the original thicknesses. The interdiffusion processes were confirmed by transmission electron microscopy, which enhanced in proportion to the annealing temperature. The strain states of the $\beta$-$(AlGa)_2O_3$ films have been analyzed showing reduced in-plane compressive strain with higher annealing temperature; and the film eventually became strain-free when the temperature was 1400 ºC corresponding to the Al composition of 77%. The proposed method is promising for the preparation of the $\beta$-$(AlGa)_2O_3$ thin films without employing sophisticated 'direct-growth' techniques for alloys.

**Introduction**

III-oxide semiconductors such as $Ga_2O_3$ have gathered rising attention recently because of their wide bandgap, excellent stability, and availability of lower-cost bulk substrates for high-performance deep-ultraviolet (DUV) solar-blind photodetectors (PDs), transparent electronics, and power electronics.[1-15] The monoclinic $β$-phase is widely recognized as the most stable crystalline structure for the III-oxides.[16] On one hand, bulk $β$-$Ga_2O_3$ crystals can be formed by methods including edge-defined film-fed growth (EFG),[17] floating zone,[18] and Czochralski.[19] On the other, thin-film deposition techniques including sol-gel solution deposition,[20] sputtering,[21] chemical vapor deposition (CVD),[22] mist CVD,[23] pulsed laser deposition (PLD),[24] molecular beam epitaxy (MBE),[25] and metalorganic chemical vapor deposition (MOCVD)[26] have been extensively employed. For further development of III-oxide optical and power devices, the progress would benefit greatly from the alloy formation and tunability of the alloy composition; and thereby the associated properties such as bandgap, absorption spectrum, and breakdown field. The alloys of interest include $(Al_xGa_{1-x})_2O_3$, $(In_xGa_{1-x})_2O_3$, and $(Al_xIn_{1-x})_2O_3$ ternaries and even quaternaries ($0 \leq x \leq 1$). In particular, the $(Al_xGa_{1-x})_2O_3$ alloys with various phases are promising candidate for shorter-wavelength and higher-power devices due to large tunable bandgaps from 4.9 ($β$-$Ga_2O_3$) to above 8.6 eV ($α$-$Al_2O_3$).

To form the $(Al_xGa_{1-x})_2O_3$ alloys, researchers have employed RF sputtering,[27] PLD,[28,29] mist-CVD,[30] MOCVD,[31] and MBE,[32, 33,34] as summarized in Table 1 which shows the highest-reported Al compositions by different techniques for different phases. These 'direct-growth' techniques offer excellent tunability of material compositions and compositional homogeneity, albeit the highest reported Al composition for the $β$-$(Al_xGa_{1-x})_2O_3$ alloys is 61% by MBE.[32] Also Zhang et al. have demonstrated $β$-$(Al_xGa_{1-x})_2O_3$ thin films with the Al composition of 52%, above which the (-201) and higher-order peaks by X-ray diffraction (XRD) were weak or invisible.[29] However, the associated costs of the direct-growth techniques are not trivial because of various requirements: targets with different compositions, high-purity precursors/sources, or demanding environments such as an ultrahigh vacuum.

In the 1990s, Fleischer and Battiston et al. discovered that thermal annealing of $Ga_2O_3$ thin films on sapphire could lead to Al diffusion into the films.[35,36] Later, Kokubun et al. found that increased temperatures from 600 to 1200 °C during heat treatment of $Ga_2O_3$ sol-gel can boost the polycrystalline $Ga_2O_3$ bandgap from 4.95 to 5.53 eV which the authors attributed to the Al diffusion.[20] Recently, Goyal et al. varied annealing temperatures of polycrystalline $Ga_2O_3$ thin films on sapphire from 600 to 1000 °C to obtain changing bandgap from 4.63 to 5.15 eV with annealing time of 24 and 36 hours.[37] They performed secondary ion mass spectrometry (SIMS) analysis which show Al diffusion into the films, though being largely inhomogeneous. Those studies hint a viable pathway of forming $(Al_xGa_{1-x})_2O_3$ ternary alloys based on $Ga_2O_3$/sapphire templates by thermal diffusion, thereby being low cost and straightforward as opposed to the direct growth techniques. However, the Al compositions were still low and inhomogeneous with polycrystalline $Ga_2O_3$ templates in those studies, which greatly hindered the application prospects. Besides, the thin film/sapphire interface has not been systematically examined, preventing an understanding of the annealing impact.

Table 1. Summary of the highest reported Al compositions of the $(Al_xGa_{1-x})_2O_3$ thin films by various direct-growth techniques.

| Techniques | Al composition range | Phase | Ref. |
|---|---|---|---|
| RF sputtering | x up to 0.059 | $\beta$-$Al_xGa_{1-x})_2O_3$ | 27 |
| PLD | x up to 0.98 | $\alpha$-$Al_xGa_{1-x})_2O_3$ | 29 |
| Mist-CVD | x up to 0.81 | $\alpha$-$Al_xGa_{1-x})_2O_3$ | 30 |
| MOCVD | x up to 0.40 | $\beta$-$Al_xGa_{1-x})_2O_3$ | 31 |
|  | y up to 0.39 | $\gamma$-$Al_yGa_{1-y})_2O_3$ |  |
| MBE | x up to 0.61 | $\beta$-$Al_xGa_{1-x})_2O_3$ | 32 |

In this work, we have employed $\beta$-$Ga_2O_3$/sapphire templates to obtain ternary $\beta$-$(Al_xGa_{1-x})_2O_3$ alloys by thermal annealing at various elevated temperatures. We found that controlling annealing temperature from 1000 to 1400 °C can tune the diffusion rate, leading to Al compositions from 0 up to 77%, corresponding to a large bandgap from 4.9 to 6.4 eV for $\beta$-$(Al_xGa_{1-x})_2O_3$. The corresponding material composition, film thickness, and strain state were characterized systematically. Besides, the thin film/sapphire interface was investigated by high-resolution transmission electron microscopy (HR-TEM).

**Experiment**

In this study, 50-nm thick $\beta$-$Ga_2O_3$ thin-film templates were deposited on *c*-plane sapphire substrates by the PLD. Although the PLD was utilized, it is important to note that the same studies can be performed for $\beta$-$Ga_2O_3$ thin films on sapphire grown by other techniques such as CVD,

MBE, and MOCVD because we expect that the thermal diffusion could occur regardless of the growth techniques. The PLD condition includes 800 °C heater temperature, 4.5 mTorr chamber pressure, and KrF (248 nm) excimer laser power of 400 mJ with a 5 Hz repetition rate. A single undoped $Ga_2O_3$ target provided by Sigma-Aldrich was in use. The thermal annealing was conducted in an MTI KSL-1700X-A4-DC furnace at different temperatures of 1000−1400 °C with the step of 100 °C in the air at atmospheric pressure for three hours. The crystal structures, material quality, material composition, optical properties, thin-film thickness, and interface were characterized by using XRD, SIMS, optical transmission, and TEM. The XRD 2θ scans were carried out by a Bruker D2 PHASER system with a wavelength of λ is 1.5406 Å. SIMS experiments including element depth profiling were performed by a Dynamic SIMS system from Hiden Analytical. The optical transmission was measured by a Shimadzu UV-3600 spectrophotometer. Afterward, the optical bandgap was deduced from the Tauc plot from transmission spectra. The TEM specimens were obtained by a Helios focused ion beam (FIB) system. The cross-sectional TEM images and Energy-dispersive X-ray spectroscopy (EDX) mapping were acquired by an FEI Titan ST and Gatan EDX system.

**Results and Discussion**

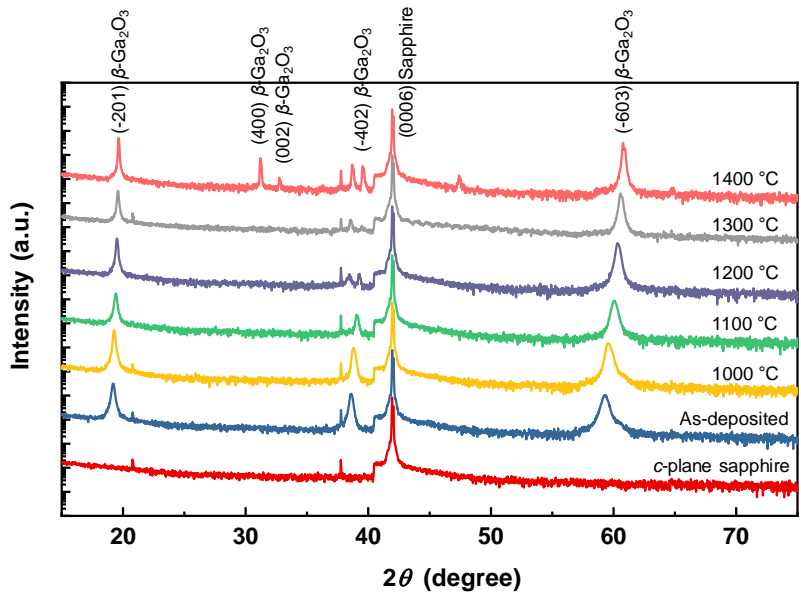

Figure 1. XRD 2θ scan spectra of the as-deposited and annealed $Ga_2O_3$ thin films on sapphire with different annealing temperatures, and a bare sapphire substrate.

Figure 1 shows the XRD 2θ scan spectra of the as-deposited and annealed $Ga_2O_3$ thin films on sapphire with different annealing temperatures for three hours, and a bare sapphire substrate. For the as-deposited sample, there are three dominating peaks from the (-201) and higher-order diffraction planes of β-$Ga_2O_3$ thin films, demonstrating single crystallinity. The (0006) sapphire substrate diffraction plane peak can also be seen. As the annealing was introduced with temperatures increasing from 1000 to 1400 °C, the (-201) peak positions shifted toward higher angles as shown in Figure 1. Table 2 summarizes the peak positions of different samples, indicating decreased out-of-plane lattice constant according to the Bragg's Law for samples annealed at higher temperatures. The annealed samples were single crystalline with the temperatures of 1000 and 1100 °C. But with the temperatures of 1200, 1300, and 1400 °C, the (-402) peak split into two peaks: one remained as the (-402) peak and the other one might belong to the (-401) diffraction plane peak. Two extra peaks (400) and (002) were observed in the 1400 °C annealed sample. These two peaks could be attributed to the thermal-strain induced by high-temperature annealing.[38]

Table 2. XRD 2θ scan peak positions of different diffraction planes extracted from Figure 1. The (-201) peak shifts of samples with respect to the as-deposited β-$Ga_2O_3$/sapphire sample are listed as the final column.

| Annealing temperature | (-201) peak | (-402) peak | (-603) peak | (0006) sapphire peak | (-201) ΔPeak shift |
|---|---|---|---|---|---|
| **As-deposited** | 18.95° | 38.36° | 59.04° | 41.68° | 0° (reference) |
| **1000 °C** | 18.98° | 38.52° | 59.28° | 41.68° | +0.03° |
| **1100 °C** | 19.13° | 38.83° | 59.80° | 41.68° | +0.18° |
| **1200 °C** | 19.25° | (-401) 38.19°/ 39.02° | 60.03° | 41.68° | +0.30° |
| **1300 °C** | 19.33° | (-401) 38.30°/ 39.20° | 60.29° | 41.68° | +0.38° |
| **1400 °C** | 19.37° | (-401) 38.43°/ 39.28° | 60.51° | 41.68° | +0.42° |

To corroborate the findings of the XRD 2θ scans indicating reduced lattice constants with higher annealing temperatures, the SIMS analysis was carried out with depth profiling of Al, Ga, and O elements for the as-deposited and annealed samples at 1000, 1200, and 1400 °C shown in Figure 2. Figure 2(a) shows that the Al concentration of the thin film is at the detection limit level indicating minimal Al diffusion. Also, the $Ga_2O_3$ film thickness is 50 nm consistent with the design of the PLD experiments. With three hours of annealing at 1000 °C, significant Al diffusion occurred unambiguously shown in Figure 2(b). The annealing also increased the film thickness to 110 nm indicating the interdiffusion process of Ga and Al atoms. However, there is a small gradient of Ga and Al concentrations throughout the film and thus the alloy material composition is not strictly homogeneous. With higher temperatures of 1200 and 1400 °C, the film thickness increased further to 190 and 250 nm with three hours of annealing, respectively, suggesting stronger interdiffusion shown in Figure 2(c)-(d). More importantly, the Al compositions are highly homogeneous throughout the entire film. All the carbon signals in Figure 2(a)-(d) are at a detection-limit level or lower, indicating and low impurity level and thus low contamination amid the process.

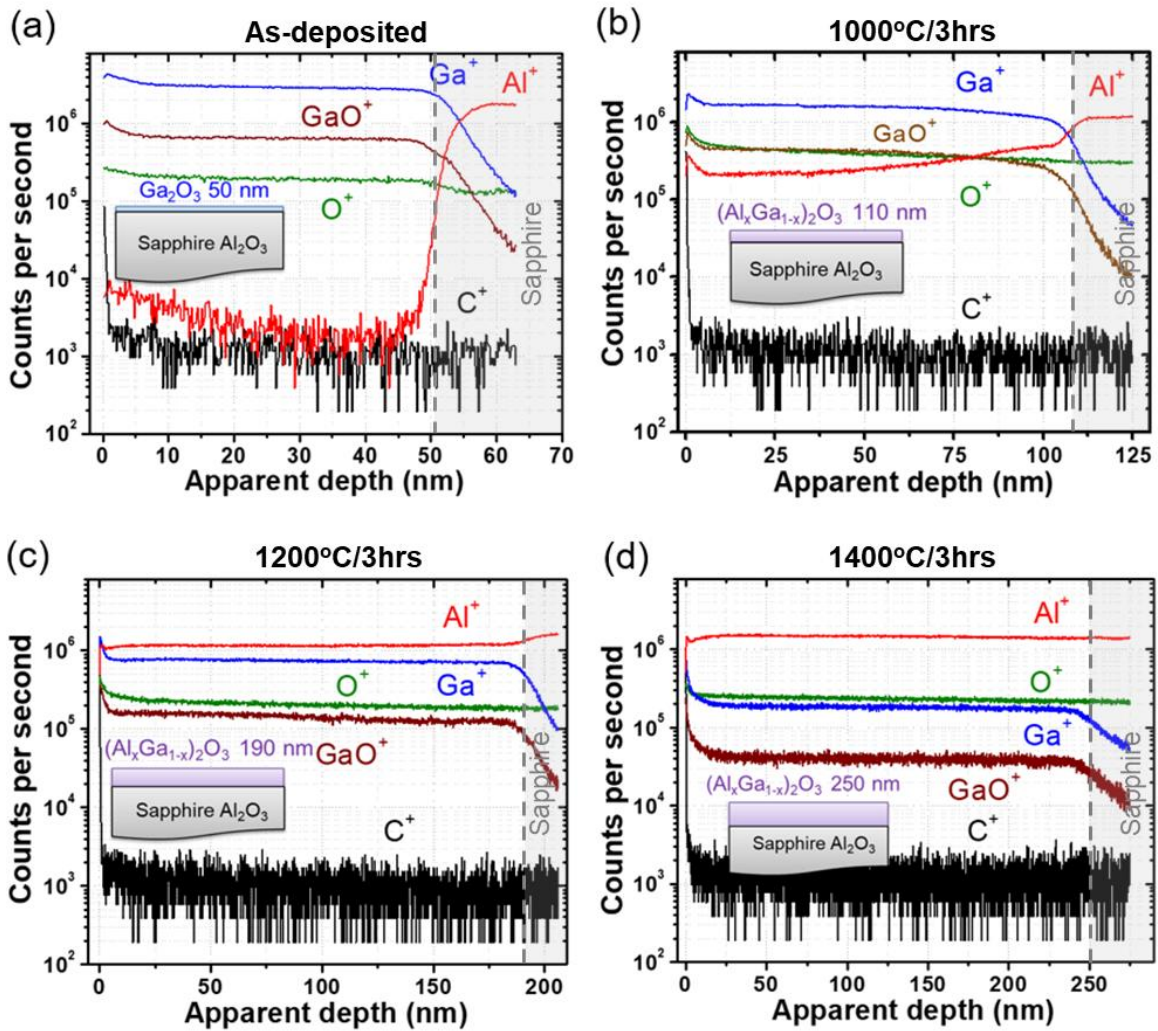

Figure 2. The SIMS depth profiles of (a) as-deposited $\beta$-$Ga_2O_3$; and samples annealed at (b) 1000 °C, (c) 1200 °C, and (d) 1400 °C. The inset figures are the sample structures with the estimated thin film thickness of each sample. The grey dash lines represent the thin film/sapphire interface.

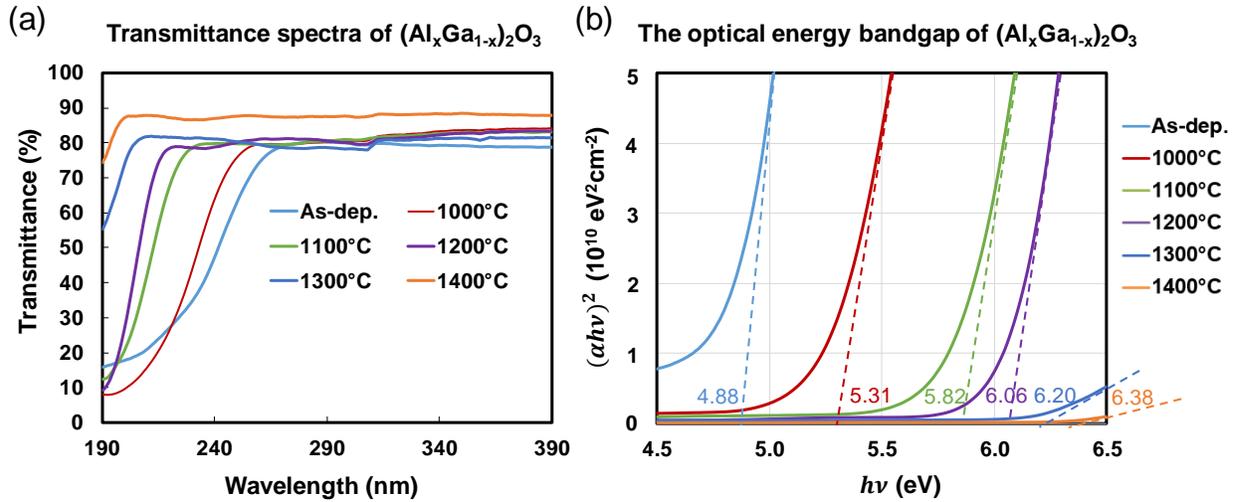

Figure 3. (a) Transmission spectra of the as-deposited β-Ga$_2$O$_3$ and the samples annealed at different annealing temperatures. (b) The $(\alpha h\nu)^2$ versus $h\nu$ plots (Tauc plot) of the β-(AlGa)$_2$O$_3$ films and the optical energy bandgap extrapolations.

To quantify the alloying and the Al composition of the annealed samples, the optical transmission measurement was conducted to obtain optical bandgap shown in Figure 3(a). Blue-shift in the absorption edge with increased annealing temperature can be observed indicating that the successful alloying occurred after the Al diffusion, which formed β-(Al$_x$Ga$_{1-x}$)$_2$O$_3$ alloy with increased Al composition. Because of the detection wavelength limitation of the spectrophotometer, we only can measure the wavelength down to 190 nm, which makes the transmission curves of high temperature (1300 °C and above) uncompletedly measured. The bandgap of annealed thin films was firstly deduced from the Tauc plot shown in Figure 3(b) calculated from the transmission spectra. Then the bandgap was determined by extrapolating linear regions of these curves to the horizontal axis. An increase in bandgap with annealing temperature was observed. The bandgaps for the as-deposited β-Ga$_2$O$_3$ film were 4.88 eV and for the three-hour annealed β-(Al$_x$Ga$_{1-x}$)$_2$O$_3$ films at 1000, 1100, 1200, 1300, and 1400 °C were 5.31, 5.82, 6.06, 6.20 and 6.38 eV, respectively.

Subsequently, the Al composition of the β-(Al$_x$Ga$_{1-x}$)$_2$O$_3$ thin films can be determined according to the equation $E_g(x) = (1-x) E_g(β\text{-Ga}_2O_3) + x E_g(θ\text{-Al}_2O_3) - b\, x\, (1-x)$, where the bandgaps of β-Ga$_2$O$_3$ (monoclinic) and θ-Al$_2$O$_3$ (monoclinic) were 4.87 and 7.24 eV and the bowing parameter b = 1.78 eV was used for the monoclinic phase.[39] Consequently, the Al compositions of the annealed β-(AlGa)$_2$O$_3$ films at 1000, 1100, 1200, 1300, and 1400 °C are 36,

58, 67, 72, and 77%, respectively. The relationship between the temperature and the Al composition shows quasi-linear behavior exhibited in Figure 4(a), suggesting that the annealing temperature is an excellent knob to control the alloy composition. Figure 4(b) shows the bandgap versus the Al composition. The results are consistent with those of the samples grown by PLD employing different $(Al_xGa_{1-x})_2O_3$ targets. [28,29,33]

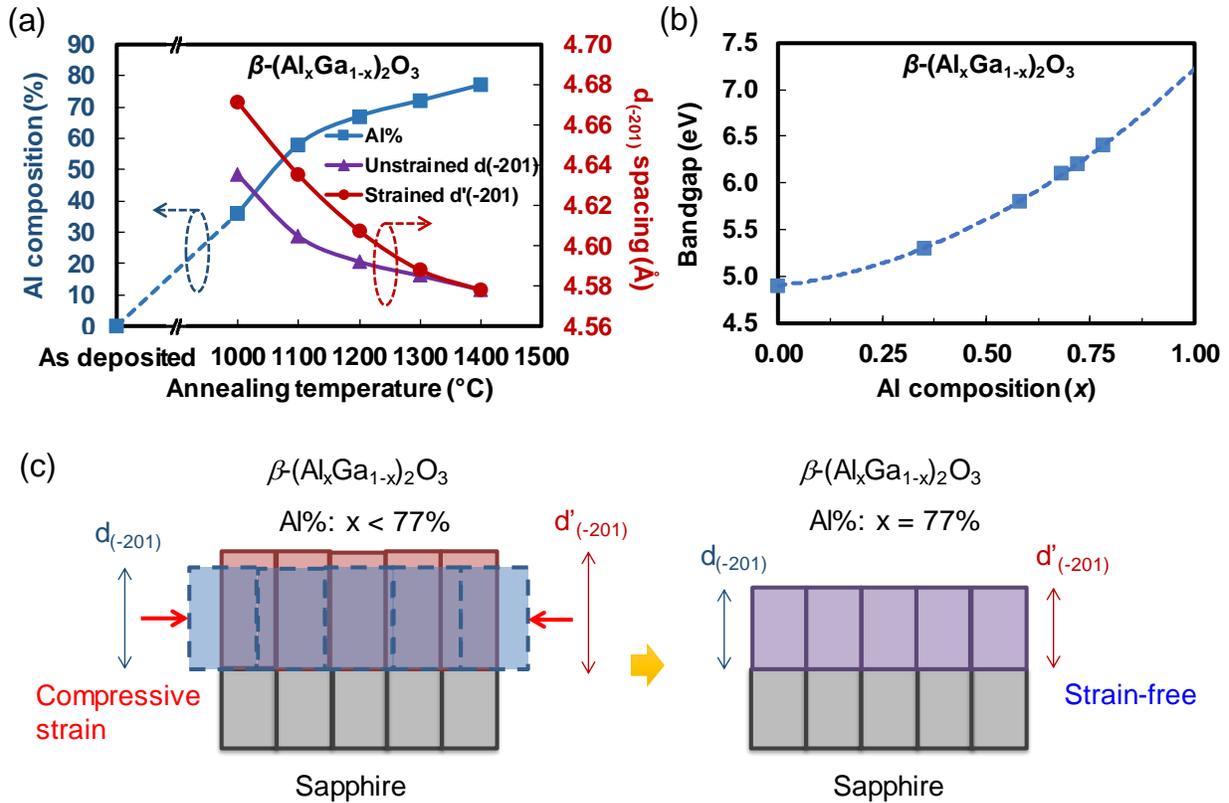

Figure 4. (a) Al composition evaluation and the calculated d-spacing of (-201) diffraction plane of the $β$-$(Al_xGa_{1-x})_2O_3$ films as a function of annealing temperature. (b) The bandgap of the $β$-$(Al_xGa_{1-x})_2O_3$ film as a function of Al composition. (c) The schematic diagrams of strain states in the $β$-$(Al_xGa_{1-x})_2O_3$ film at different Al contents.

The SIMS and transmission experiments corroborated the XRD 2θ scan results well. The shift in the XRD peaks related to the monolithic phase is caused by the Al diffusion, resulting in the $β$-$(Al_xGa_{1-x})_2O_3$ ternary alloys with larger and tunable bandgap versus the annealing temperature with the Al composition at or even higher than 77%. Previously, Peelaers *et al.* predicted that the monoclinic phase is the preferred structure for up to 71% Al incorporation into $β$-$(Al_xGa_{1-x})_2O_3$.[39] To understand this discrepancy, we performed the $β$-(AlGa)$_2$O$_3$ lattice parameters calculated from the Al content of these different temperatures annealed samples in Table 3. The $\beta$-(Al$_x$Ga$_{1-x}$)$_2$O$_3$ lattice parameters were obtained by follows: a = (12.21−0.42x), b = (3.04−0.13x), c = (95.81−0.17x), and β = (103.87+0.31x), which follow the Vegard's law and Kranert et al.[40] After that, from the d-spacing of the (*hkl*) plane of the monoclinic structure in Equation (1), we can calculate the d-spacing of (-201) diffraction planes of $\beta$-(AlGa)$_2$O$_3$ for each Al content (annealing temperature). Such a d-spacing evaluation method from given Al compositions determined from the bandgap in the Tauc plot (Figure 3(b)) is considered as an unstrained d$_{(-201)}$ of $\beta$-(AlGa)$_2$O$_3$ film.

$$\frac{1}{d^2} = \frac{h^2}{a^2 \sin^2\beta} + \frac{k^2}{b^2} + \frac{l^2}{c^2 \sin^2\beta} + \frac{2hl\cos\beta}{ac\sin^2\beta} \quad \text{---(1)}$$

However, one can also extract another set of lattice parameters and d-spacing values from the XRD (-201) peak positions of different annealed samples. Such a d-spacing evaluation method from the experimental data of the (-201) peak position determined by the Bragg's Law 2dsinθ = nλ and Equation (1) is considered as a strained d'$_{(-201)}$ of $\beta$-(AlGa)$_2$O$_3$ film on *c*-plane sapphire because of its direct measurement the d-spacing of the (-201) plane in the strained sample. The calculated strained and unstrained d-spacing of the (-201) plane of the $\beta$-(Al$_x$Ga$_{1-x}$)$_2$O$_3$ films as a function of temperature are shown in Figure 4(a).

Table 3. The $\beta$-(AlGa)$_2$O$_3$ lattice parameters calculated through the Al% content (evaluated from Tauc plot E$_g$) of the annealed samples. The d-spacing of the (-201) diffraction planes extracted from different methods, the Al% content and XRD 2θ scan (-201) peaks, and their differences are listed in the table.

| | Annealing temperature | 1000 °C | 1100 °C | 1200 °C | 1300 °C | 1400 °C |
|---|---|---|---|---|---|---|
| **Evaluated from bandgap (Tauc Plot)** | Al content | 36 % | 58 % | 67 % | 72 % | 77 % |
| | a (Å) | 12.06 | 11.97 | 11.93 | 11.91 | 11.89 |
| | b (Å) | 2.99 | 2.96 | 2.95 | 2.95 | 2.94 |
| | c (Å) | 5.75 | 5.71 | 5.70 | 5.69 | 5.68 |
| | β | 103.98° | 104.05° | 104.08° | 104.09° | 104.11° |
| | Unstrained d$_{(-201)}$ (Å) | 4.635 | 4.605 | 4.592 | 4.585 | 4.578 |
| **Extracted from XRD 2θ-scan** | (-201) peak | 18.98° | 19.13° | 19.25° | 19.33° | 19.37° |
| | Strained d'$_{(-201)}$ (Å) | 4.671 | 4.635 | 4.608 | 4.588 | 4.578 |
| **Difference** | Δd$_{(-201)}$ (Å) | 0.036 | 0.031 | 0.015 | 0.003 | 0.000 |
| | (d'$_{(-201)}$-d$_{(-201)}$) / d$_{(-201)}$ | 0.78 % | 0.66 % | 0.33 % | 0.06 % | 0.00 % |

The differences between the strained and unstrained d-spacing, i.e. $\Delta d_{(-201)} = d'_{(-201)} - d_{(-201)}$ are listed in the Table 3. With the higher annealing temperature and thus higher Al content, the strained and unstrained d-spacing difference ($\Delta d_{(-201)}$) becomes smaller. It becomes zero when the annealing temperature raised to 1400 °C (Al%=77%). Figure 4(c) shows the schematic diagrams of strain states in the $\beta$-(AlGa)$_2$O$_3$ film on a sapphire substrate at different Al contents. When $\beta$-(AlGa)$_2$O$_3$ film formed on top of sapphire, it suffered from compressive strain in the in-plane direction resulting in larger strained d'$_{(-201)}$. However, with higher annealing temperature, more Al diffused into the $\beta$-(AlGa)$_2$O$_3$ film to result in increased Al content increase but decreased lattice constant, thereby reducing the in-plane compressive strain. When the Al content reached 77% (1400 °C), the strained d'$_{(-201)}$ and unstrained d$_{(-201)}$ become identical which shows strain-free state. This result shows the in-plane compressive strain existed in the $\beta$-(AlGa)$_2$O$_3$ film until the Al content reached the strain-free condition, which is 77% in this study. Therefore, this study demonstrates higher Al composition than Peelaers' theoretical prediction without showing any $\beta$-(Al$_x$Ga$_{1-x}$)$_2$O$_3$ and $\alpha$-(Al$_x$Ga$_{1-x}$)$_2$O$_3$ mixed-phase phenomenon.[41]

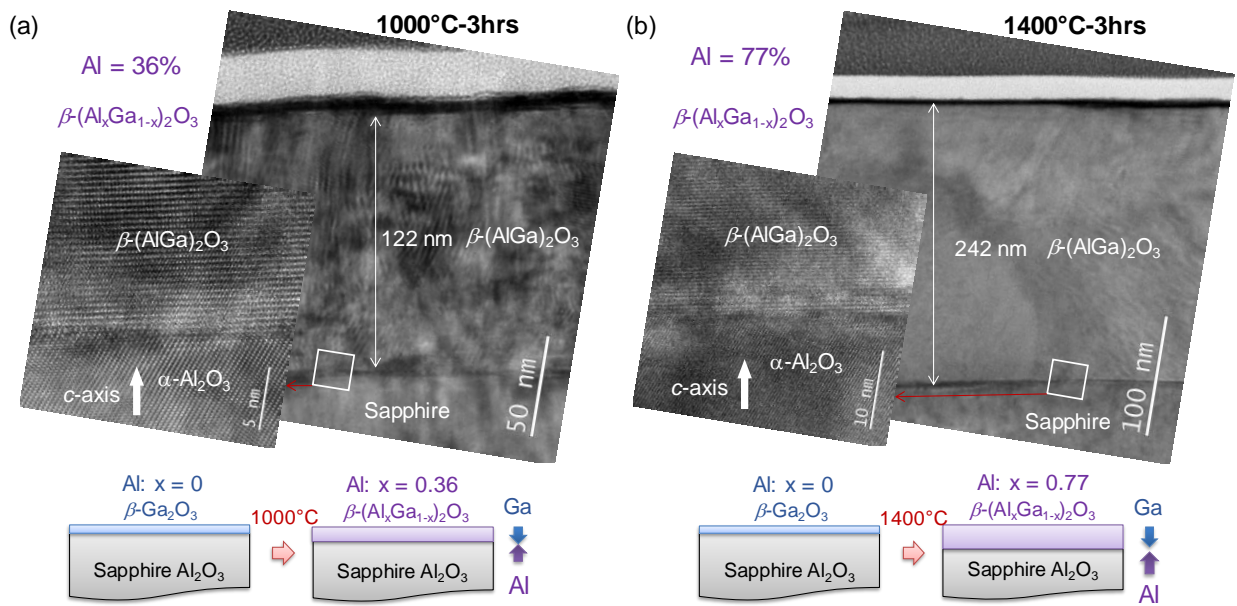

Figure 5. Cross-sectional TEM images of the samples annealed at (a) 1000 °C and (b) 1400 °C for three hours. The inset HR-TEM images in (a) and (b) show a transition layer between the $\beta$-(AlGa)$_2$O$_3$ thin film and sapphire. The schematic diagrams show the interdiffusion process during the high-temperature annealing.

The SIMS results in Figure 2 indicated that the interdiffusion of Al and Ga atoms caused thicker β-($Al_xGa_{1-x}$)$_2$O$_3$ thin films at higher annealing temperatures. Therefore, the understanding of the β-($Al_xGa_{1-x}$)$_2$O$_3$/sapphire interface is crucial. To investigate the interface, we have performed cross-sectional TEM experiments of the samples annealed at 1000 and 1400 °C for three hours. Figure 5(a) shows the TEM images of the β-(AlGa)$_2$O$_3$ film annealed at 1000 °C corresponding to the 36% Al composition. The thickness of the β-($Al_{0.36}Ga_{0.64}$)$_2$O$_3$ thin film was about 122 nm thick which in good agreement with the thickness that the SIMS profiling estimated (Figure 2(b)). Figure 5(b) shows the TEM images of the 1400 °C-annealed sample with the 77% Al composition. The film thickness of β-($Al_{0.77}Ga_{0.23}$)$_2$O$_3$ was about 242 nm which agreed with the SIMS result (Figure 2(d)). According to the insets of Figure 5(a) and (b), there was a transition layer between the annealed β-(AlGa)$_2$O$_3$ thin film and sapphire substrate. The schematic diagrams show the Al and Ga interdiffusion process result in thicker film with increased annealing temperature.

Cross-sectional HR-TEM image of the 1000 and 1400 °C annealed samples in the vicinity of the transition layers are shown in Figure 6(a) and (c). The transition layers around 5 nm thick between the β-AlGa$_2$O$_3$ film and the sapphire substrate can be identified for both samples. The fast Fourier transform (FFT) diffraction patterns of both annealed samples confirm the α-phase corundum crystal structure for the sapphire substrate and the monoclinic β-phase crystal structure for the β-(AlGa)$_2$O$_3$ thin films; and the FFT diffraction patterns of the transition layers of both samples show the mixed crystal structures of α-phase and β-phase. The STEM HAADF images and EDX maps of the two samples in the vicinity of the transition layers are shown in Figure 6(b) and (d). The Al and Ga gradients in the EDX maps show that the Al atoms diffused from the sapphire substrate upward to the (AlGa)$_2$O$_3$ film layer, which was the opposite of the Ga diffusion. Meanwhile, the O distribution was homogeneous. In Figure 6(b) and (d), the EDX signal overlap maps show a tiny orange-ish region comprising Al red and Ga green signals at the interface of β-(AlGa)$_2$O$_3$/sapphire reaffirming the existence of the transition layer. For the EDX spectra in Figure 6(b) and (d), the Al signal intensity is weaker than that of Ga for the sample annealed at 1000 °C in (b); and the Al signal intensity is stronger than that of Ga for the sample annealed at 1400 °C in (d). The results reiterate that with higher temperature annealing, more Al atoms diffused into the AlGa$_2$O$_3$ layer. It is noted that because the film thickness increased amid the annealing due to the interdiffusion process, one could design experiments with different initial thicknesses to reach the desired thicknesses and compositions by utilizing the annealing temperature knob.

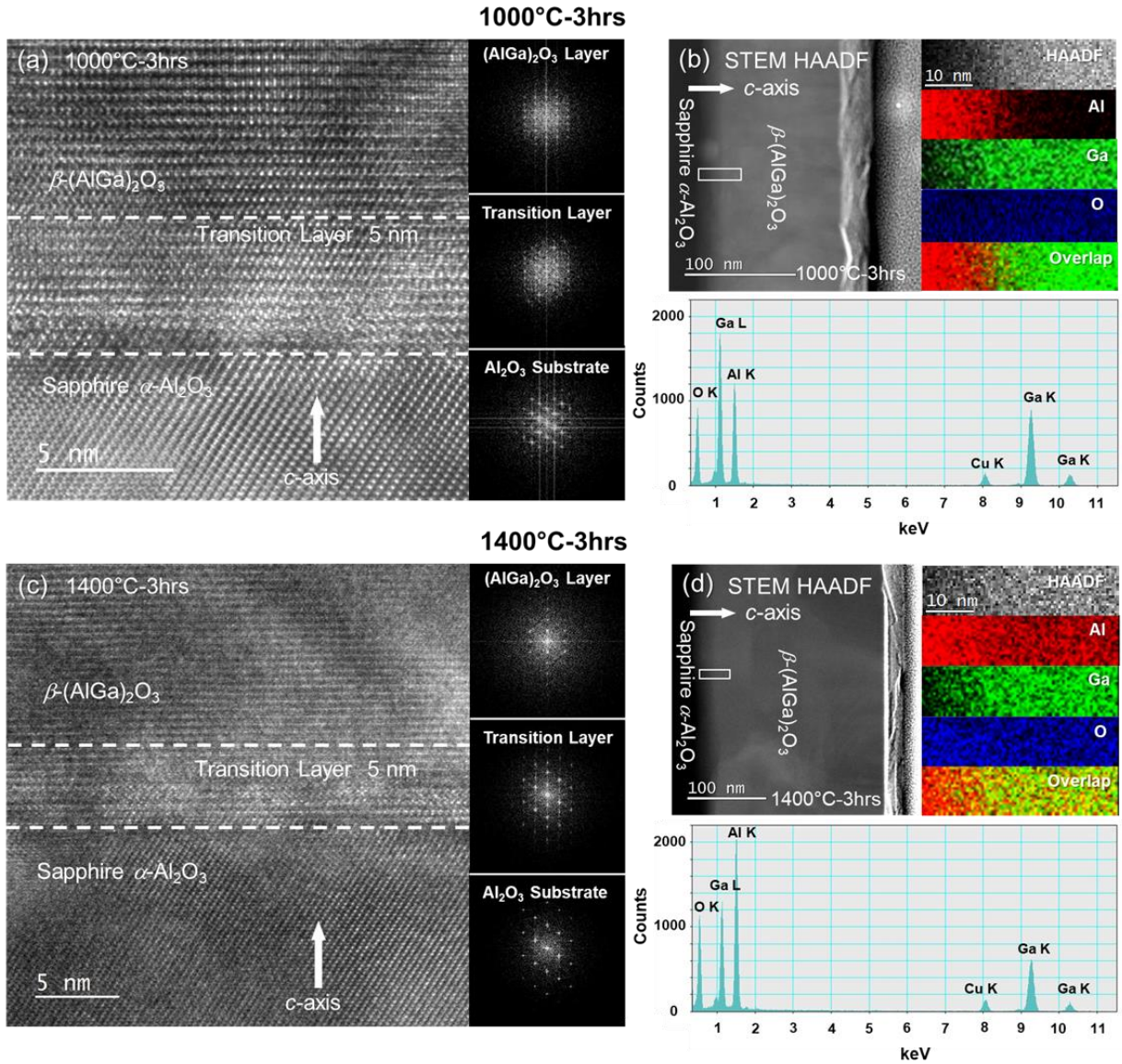

Figure 6. Cross-sectional HR-TEM images and the FFT diffraction patterns in the vicinity of the transition layers of the β-(AlGa)$_2$O$_3$ thin films annealed at (a) 1000 ºC and (c) 1400 ºC. The corresponding HAADF images, EDX maps with elements of interest (Al, Ga, O, and overlapped) and EDX spectra in the vicinity of the transition region (white rectangular area) are shown in (b) and (d), respectively.

## Conclusion

Based on the binary $\beta$-$Ga_2O_3$ templates on sapphire substrates, we have employed the thermal annealing method to demonstrate $\beta$-$(Al_xGa_{1-x})_2O_3$ ternary thin films with Al compositions from 0 to record-high 77%, corresponding to optical bandgaps from 4.9 to 6.4 eV. The ternary alloy formation was caused by the interdiffusion of Ga and Al atoms in the template and substrate, respectively, which was found to be proportional to the annealing temperatures between 1000 and 1400 °C. Thus, the annealing temperatures were found to be an excellent knob to control the Al composition. The interdiffusion was confirmed by the TEM experiments. The Al compositions of the $\beta$-$(Al_xGa_{1-x})_2O_3$ ternary thin films were largely homogeneous. With lower annealing temperatures, the $\beta$-$(AlGa)_2O_3$ thin films were under in-plane compressive strain, which decreased with higher temperatures and eventually reduced to zero at 1400 °C. The proposed method does not involve direct-growth techniques for alloys and therefore is promising for straightforward and low-cost production of the ternary $\beta$-$(AlGa)_2O_3$ alloys with desired Al compositions and thicknesses for deep UV and power devices.

## Acknowledgment:

The KAUST authors would like to acknowledge the support of appreciating the support of KAUST Baseline BAS/1/1664-01-01, GCC Research Council REP/1/3189-01-01, and Competitive Research Grants URF/1/3437-01-01 and URF/1/3771-01-01.